\documentclass{raa}           
\usepackage{graphicx,times}
\usepackage{natbib}
\usepackage{amssymb,amsmath}
\bibpunct{(}{)}{;}{a}{}{,}

\usepackage[a4paper=true,pagebackref=false]{hyperref}
\hypersetup{pdftitle = Shape Parameters of 1991 to 2016 Solar Corona, pdfauthor = Rhorom Priyatikanto, pdfsubject= Solar Corona, pdfkeywords = solarcorona solaractivity} 
\hypersetup{colorlinks = true, linkcolor = green, anchorcolor = red, citecolor = blue, filecolor = red, pagecolor = red, urlcolor = red}
\usepackage{breakurl}

\newcommand{\drj}{^{\circ}}
\newcommand{\rsun}{R$_{\odot}$}
\newcommand{\rmax}{R_{\text{max}}}

\begin{document}

\title{Shape Parameters of 1991 to 2016 Solar Corona}
\volnopage{ {\bf 2012} Vol.\ {\bf X} No. {\bf XX}, 000--000}
   \setcounter{page}{1}

\author{Rhorom Priyatikanto\inst{1,2}}
\institute{
Astronomy Program, Faculty of Mathematics and Natural Sciences, Institut Teknologi Bandung, Bandung 40132, Indonesia\\
        \and
Space Science Center, National Institute of Aeronautics and Space, Bandung 40173, Indonesia; {\it rhorom.priyatikanto@lapan.go.id}\\
\vs \no
}

\abstract{
The global structure of solar corona observed in optical window is governed by the global magnetic field with different characteristics over solar activity cycle. Ludendorff flattening index becomes a popular measure of the global structure of solar corona as observed during eclipse. In this study, 15 digital images of solar corona from 1991 to 2016 were analyzed in order to construct the coronal flattening profiles as a function of radius. In most of the cases, the profile can be modeled with 2nd order polynomial function so that the radius with maximum flattening index ($\rmax$) can be determined. Along with this value, Ludendorff index ($a+b$) was also calculated. Both Ludendorff index and $\rmax$ show anti-correlation with monthly sunspot number, though the $\rmax$ values are more scattered. The variation of $\rmax$ can be regarded as the impact of changing coronal brightness profile over equator.
}

\authorrunning{R. Priyatikanto}
\titlerunning{Shape Parameters of 1991 to 2016 Solar Corona}
\maketitle

\section{Introduction}
\label{sect:intro}

Corona is the outer part of solar atmosphere with density of $\sim10^{15}$ m$^{-3}$ and temperature of millions of Kelvin. This layer has a total brightness of about $4\times10^{-6}$ times to the brightness of solar phototsphere \citep{hanaoka2012} such that observable only in short wavelength range (extreme ultraviolet and x-ray) or when the glaring photosphere is blocked by coronagraph or lunar disk during solar eclipse. Therefore, relatively rare occurrence of total solar eclipses provide opportunity to study solar corona in optical window (or white light) from the ground. In this window, the corona can be categorized into K (\emph{kontinuerlich}) corona and F (\emph{Fraunhofer}) corona with different properties. Continuum radiation of K corona, which dominates the inner part ($r<2$ \rsun), is caused by Thompson scattering of photosphere radiation by electron in the corona. In the outer part, interplanetary dust scatter Sun's radiation and create the F corona \citep{foukal2004}.

Global structure of solar corona observed in optical window represents the electron distribution in this layer which is influenced by local and global magnetic field extending from photosphere to corona \citep[e.g.][]{sykora2003,pasachoff2009}. The change of coronal structure or shape over solar activity cycle is clearly observed. During minimum, there are few active regions and helmet streamers are relatively concentrated near the equator so that the corona tends to be flattened. Conversely, solar corona becomes more radially symmetric during maximum phase as the streamers are more evenly distributed over heliographic latitudes.

Quantitative parameters were defined to describe global structure of solar corona, some of which are photometric or Ludendorff flattening index \citep{ludendorff1928}, geometric flattening index \citep{nikolsky1956}, angular extent of streamer-free polar regions \citep{loucif1989}, modified flattening index \citep[accounting magnetic tilt][]{gulyaev1997}, and latitudinal span of helmet streamers \citep{tlatov2010}. Among those parameters, Ludendorff index becomes the most popular measure and regularly obtained from every corona observation during solar eclipse \citep{pishkalo2011}. This index can be regarded as the flattening of solar corona at 2{\rsun} heliocentric distance whose values range from $\sim0$ during solar maximum to $\sim0.4$ during solar minimum.

\cite{pishkalo2011} has already compiled Ludendorff index from 1851 to 2010 and demonstrated the correlation between this index and monthly sunspot number ($SSN$). However, a rather scattered data obscures this correlation. Similar problem occurred at the variation of flattening index as a function of solar activity phase. From particular eclipse event, some observers may get different flattening index. This difference arise from several influencing factors such as observational bias \citep{sykora1999}, diverse detector characteristics (e.g. film emulsion), exposure time, number of isophotal contours used to calculate flattening index, poorly-oriented image, to the different statistics implemented \citep{pishkalo2011}. These factors can be minimized or even eliminated by implementing homogeneous method of analysis that becomes a main focus of this study.

The objectives of this study are to re-analyze publicly available coronal images from 1991 to 2016 solar eclipses and to construct radial profiles of flattening index. From each profile, $\rmax$ that represents the equatorial radius with maximum flattening index can be determined so that the variation of this value over solar cycle can be examined together with Ludendorff index. Data used and method applied in this study are explained in \autoref{sec:data}, while the result and discussion are presented in \autoref{sec:result}. The study is concluded in \autoref{sec:conclusion}.

\section{Data and Method}
\label{sec:data}

For present study, fifteen 8-bits solar coronal images with \texttt{.jpg} extension taken during total solar eclipses 1991 to 2016 were compiled from various sources. Starting year of 1991 was chosen because more observers both amateurs and professionals started to share their images in online media. List of total solar eclipse was obtained from \emph{Fifty Year Canon Solar Eclipses, 1986-2015} \citep{espenak1990}. In this 26 years time span, there are 17 total solar eclipses (excluding hybrids), but no appropriate coronal images for 30 June 1992 and 23 November 2003 found. The totality of these two eclipses crossed over desolate areas.

\begin{figure}
\centering
\includegraphics[width=0.7\textwidth]{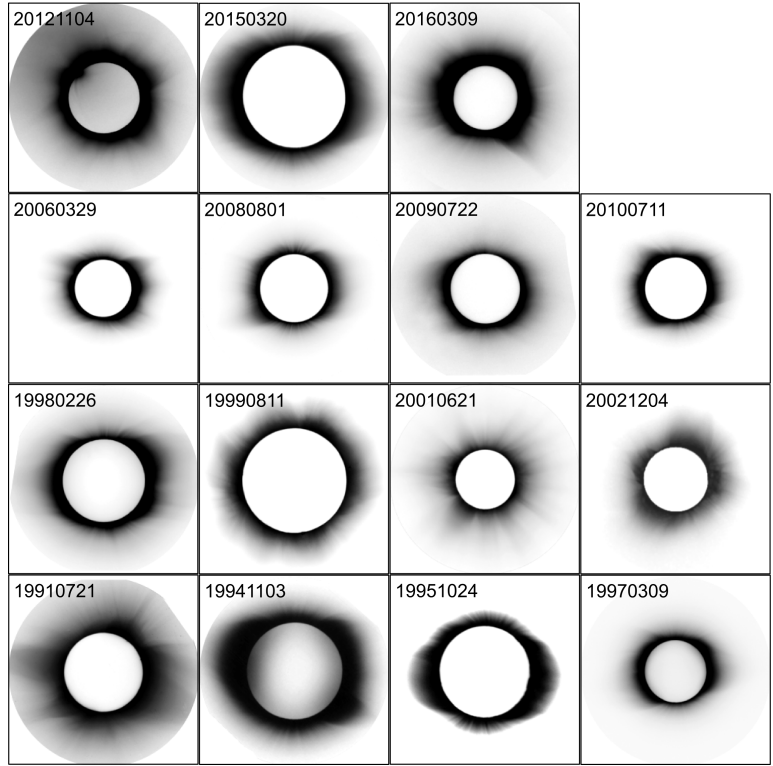}
\caption{Inverted images of solar corona which were obtained during 1991 to 2016 solar eclipses. Each image was crop into circular shape so that image rotation is easier.}
\label{fig1}
\end{figure}

Beside coronal image from the last eclipse (9 March 2016), the images were downloaded from online publication media with the help of browsing machine. \autoref{fig1} displays the inverted images while \autoref{tab1} summarizes informations related to the images. To ensure the data homogeneity, some criteria were implemented in image selection. First, the coronal image should be grabbed using digital camera equipped with neutral density filter. Image obtained using radially graded filter was not selected as it enhances outer part of solar corona and changes the brightness profile. Second criteria concerns observational field of view that is localized around 2 and 3 times Sun's angular diameter. With this condition, pixel resolution is sufficiently good while the following analysis can be conducted to the outer corona. Saturated images were obviously discarded.

\begin{table}
\centering
\caption{Some informations related to the coronal images used in this study.}
\label{tab1}
\renewcommand{\tabcolsep}{0.15cm}
\renewcommand{\arraystretch}{1.2}

\begin{tabular}{lp{2.2cm}p{1.8cm}p{5cm}p{2.5cm}}
\hline
Date & Obs. Site & Observer & Source & Reference Image \\
\hline\hline
1991-07-11  & Baja California, Mexico & M.A. Stecker & \url{http://mstecker.com/pages/astse_anr28se1b.htm} & \cite{sykora1999} \\\hline
1994-11-03  & Chile & M. Mobberley & \url{http://martinmobberley.co.uk/TSE.html} & \cite{badalyan2008} \\\hline
1995-10-24  & Ghanoli, India & G. Schneider & \url{http://nicmosis.as.arizona.edu:8000/ECLIPSE_WEB/ECLIPSE_95/UMBRAPHILE_DEBUT_1995.html} & \cite{rusin1996} \\\hline
1997-03-09  & Chita, Russia &  & \url{https://commons.wikimedia.org/w/index.php?curid=24398440} & \cite{pinter1997} \\\hline
1998-02-26  & Aruba & C.J. Lancaster & \url{http://carllancaster.com/eclipse.htm} & \cite{dorotovic1999} \\\hline
1999-08-11  & Turkey & R.C. Hoagland & \url{http://yowusa.com/nostradamus/KOT_home/KOT/hoagland_rebuttal/hoagland_rebuttal.shtml} & \cite{badalyan2008} \\\hline
2001-06-21  & Lusaka, Zambia & W. Carlos & \url{http://web.williams.edu/Astronomy/eclipse/eclipse2001/2001total/} & \cite{reginald2003} \\\hline
2002-12-04  & Ceduna, Australia & J. Pasachoff et al. & \url{https://svs.gsfc.nasa.gov/cgi-bin/details.cgi?aid=2655} & \\\hline
2006-03-29  & Tokat, Turkey & K. Kulac & \url{https://upload.wikimedia.org/wikipedia/commons/2/23/Total_solar_eclipse_2006-04-29.JPG} & \cite{pasachoff2007, stoeva2008} \\\hline
2008-08-01  & Novosibirsk, Russia & M. Pozojevic & \url{http://www.hrastro.com/SolarEclipse2008_Novosibirsk/} & \cite{pasachoff2009,skomorovsky2012} \\\hline
2009-07-22  & Varanasi, India & M. Dayyala & \url{https://commons.wikimedia.org/wiki/File:Total_solar_eclipse_on_22nd_July_at_Varanasi,India.jpg} & \cite{pasachoff2011a} \\\hline
2010-07-11  & Polynesia & C. Bowden & \url{http://www.weymouthastronomy.co.uk/gallery/eclipse/eclipse.php?show=2} & \cite{pasachoff2011b} \\\hline
2012-11-13  & Australia & NCAR/HAO & \url{https://www2.ucar.edu/for-staff/update/eclipse-12-making-mini-megamovie} & \cite{pasachoff2015} \\\hline
2015-03-20  & Scotland & W.E. Macduff & \url{https://crashmacduff.files.wordpress.com/2015/03/tse.jpg} & \cite{bazin2015} \\\hline
2016-03-09  & Sigi, Indonesia & A. Rachman &  & \cite{dani2016} \\
\hline
\end{tabular}

\end{table}

\begin{figure}
\centering
\includegraphics[scale=0.6]{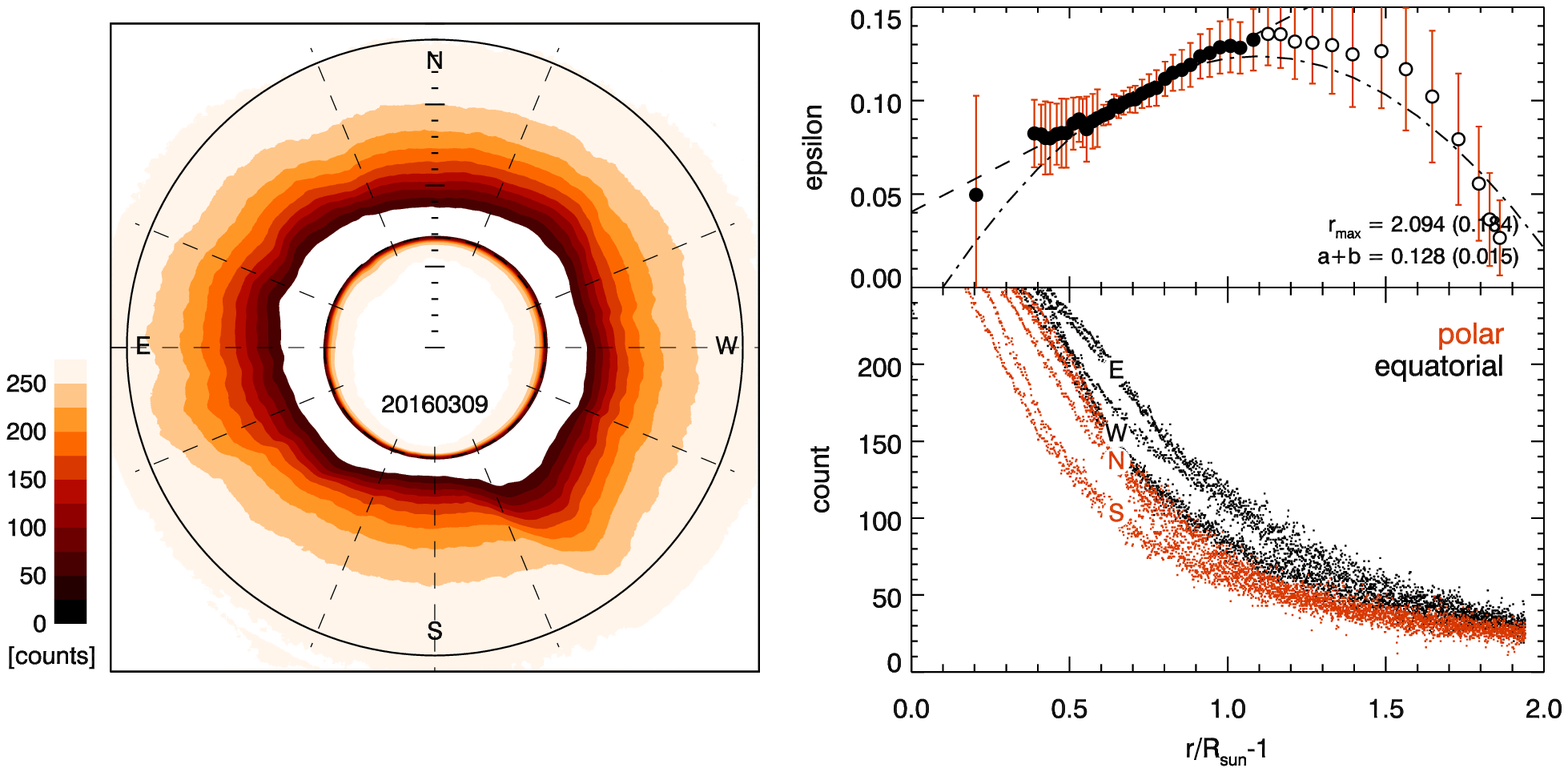}
\caption{An example of the shape analysis of coronal image taken during 9 March 2016 solar eclipse. Left panel displays brightness contours of the corona and the dashed lines mark twelve directions in which brightness profiles were extracted. Bottom right panel shows brightness profiles extracted in equatorial (dark-color) and polar (light-color) directions. Several groupings are observed due to asymmetric shape of the corona. Profiles in the north, east, south, and west directions are marked accordingly. Top right panel shows the obtained flattening indices and their errors at different equatorial radii, together with the fitted quadratic (dot-dashed) and linear (dashed) functions. In this panel, all circles represent the data used for 2nd order polynomial fitting, filled circles for linear fitting.}
\label{fig2}
\end{figure}

First step to do after downloading the images was to determine the orientation of solar disk according to the reference images published in some literatures (see \autoref{tab1}). Exact orientation of solar pole is in a great importance during the analysis of global structure of solar corona. The next process was to extract brightness profile of the corona as a function of radius (e.g. counts versus radius, see \autoref{fig2}), especially in the angle of $0\drj$, $0\drj\pm22.5\drj$, $180\drj$, and $180\drj\pm22.5\drj$ that represent equatorial directions and also $90\drj$, $90\drj\pm22.5\drj$, $270\drj$, and $270\drj\pm\drj$ that represent polar directions. The resulted profiles will be used to calculate coronal flattening index as defined by the following formula:
\begin{equation}
\epsilon\equiv\dfrac{d_0+d_1+d_2}{D_0+D_1+D_2}-1,
\end{equation}
where $d_0,d_1,d_2$ are coronal diameter in equatorial directions, while $D_0,D_1,D_2$ are measured diameter in polar directions. The diameter is just the sum of two opposite radii with specific pixel counts and it is deduced from previously constructed brightness profile. The values of $\epsilon$ range from $0.0$ to $0.4$, depend on the phase of solar activity (and some other factors). Typically, flattening index increases by radius in the inner corona (up to $\rmax=1.5-2.5$ \rsun) and than declines to minimum value. Previous authors often create isophotal contours of solar corona and calculate flattening index in each contour, while in this study, continuous brightness profile was used to determine the index in smaller interval of radius. In this way, the uncertainty of the flattening index can be calculated using simple rule of error propagation. Large dispersion in brightness profile tends to produce larger uncertainty of flattening index. At last, the resulted plot of flattening index versus equatorial radius can be regarded as flattening profile of the solar corona.

Following the typical increase and decrease of flattening index over equatorial radius ($r$), it is plausible to model the flattening profile using 2nd order polynomial function. The fitting was implemented to the data, particularly in the range of $r\leq 3$ R$_{\odot}$. This restriction is necessary since error of $\epsilon$ increases by radius or the drop of coronal brightness. Besides, statistical weighting was applied in order to get more robust fitting. Based on this fitting, the radius with maximum flattening index ($\rmax$) and its uncertainty can be obtained.

The value of $\rmax$ was used as the structural boundary between inner and outer corona. Flattening index of the inner part can be modeled using linear function:
\begin{equation}
\epsilon=a+b\left(\dfrac{r}{R_{\odot}}-1\right)
\end{equation}
and the summation of the two coefficients of regression represents Ludendorff flattening index which is defined as coronal flattening at $r=2$ R$_{\odot}$. Additionally, the uncertainty of Ludendorff index was calculated according to the uncertainties of the two coefficients.

\section{Result and Discussion}
\label{sec:result}

\subsection{Flattening Profile}
\begin{figure}
\centering
\includegraphics[scale=0.6]{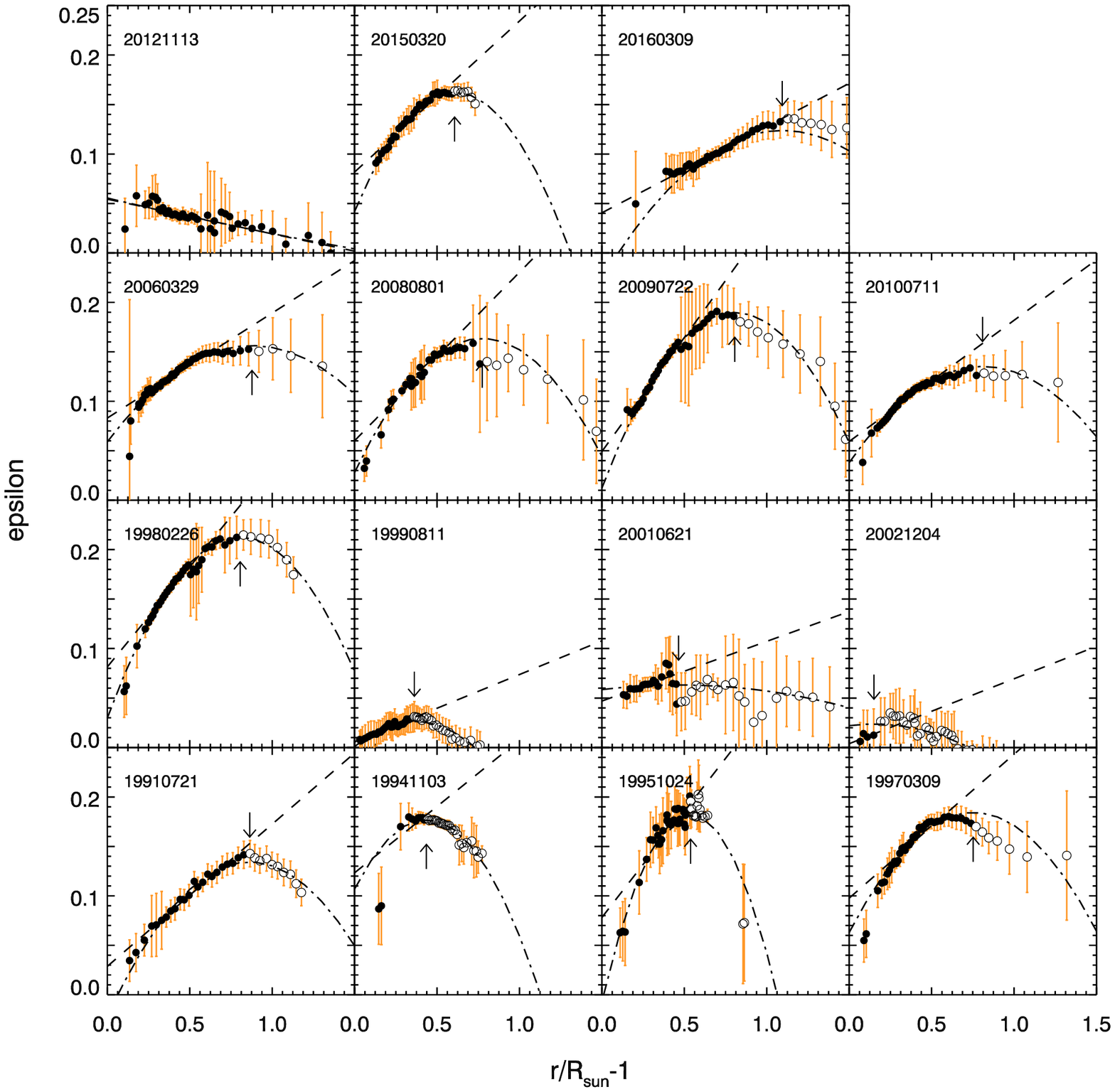}
\caption{Flattening profile ($\epsilon$ as a function of radius) produced from various coronal images. Quadratic function (dot-dashed) was fitted to the data to obtain radius with maximum $\epsilon$ ($\rmax$). Data with $r\leq\rmax$ (filled circles) were used in linear regression (dashed) to determine Ludendorff index.}
\label{fig3}
\end{figure}

\begin{table}
\centering
\caption{Summary of the shape parameters ($\rmax$ and $a+b$) obtained in this study together with their uncertainties. The corresponding phases of solar activity ($\Phi$) and flattening indices from various literatures are also presented.}
\label{tab2}
\renewcommand{\arraystretch}{1.2}

\small
\begin{tabular}{lccccc|cl}
\hline
Date & $\Phi$ & $\rmax$ & $\sigma_{R\text{max}}$ & $a+b$ & $\sigma_{a+b}$ & \multicolumn{2}{c}{$a+b$ from literatures} \\
 & & [R$_{\odot}$] & [R$_{\odot}$] & & & \\
\hline
\hline
1991-07-21  &  $-0.70$  &  $1.86$  &  $0.18$  &  $0.17$  &  $0.02$  &  $0.00$  &  \cite{sykora1999} \\
1994-11-03  &  $-0.22$  &  $1.44$  &  $0.16$  &  $0.26$  &  $0.09$  &  $0.14$  &  \cite{badalyan2008} \\
1995-10-24  &  $-0.09$  &  $1.54$  &  $0.11$  &  $0.29$  &  $0.03$  &  $0.28$  &  \cite{rusin1996} \\
1997-03-09  &  $0.22$  &  $ 1.75$  &  $0.09$  &  $0.24$  &  $0.01$  &  $0.20$  &  \cite{pinter1997} \\
1998-02-26  &  $0.48$  &  $ 1.81$  &  $0.10$  &  $0.28$  &  $0.01$  &  $0.21$  &  \cite{dorotovic1999} \\
1999-08-11  &  $0.87$  &  $ 1.36$  &  $0.12$  &  $0.07$  &  $0.02$  &  $0.04$  &  \cite{pishkalo2011} \\
2001-06-21  &  $-0.83$  &  $1.47$  &  $0.70$  &  $0.11$  &  $0.04$  &  $0.07$  &  \cite{pishkalo2011} \\
2002-12-04  &  $-0.64$  &  $1.15$  &  $0.27$  &  $0.07$  &  $0.39$  &  $0.09$  &  \cite{pishkalo2011} \\
2006-03-29  &  $-0.22$  &  $1.88$  &  $0.15$  &  $0.19$  &  $0.01$  &  $0.17$  &  \cite{pishkalo2008} \\
2008-08-01  &  $0.09$  &  $ 1.77$  &  $0.09$  &  $0.23$  &  $0.01$  &  $0.21$  &  \cite{pishkalo2009} \\
2009-07-22  &  $0.25$  &  $ 1.81$  &  $0.07$  &  $0.28$  &  $0.01$  &  $0.24$  &  \cite{pishkalo2011} \\
2010-07-11  &  $0.40$  &  $ 1.81$  &  $0.12$  &  $0.18$  &  $0.01$  &  $0.24$  &  \cite{pishkalo2011} \\
2012-11-04  &  $0.77$  &  $ 1.30$  &  $0.50$  &  $0.02$  &  $0.01$  &  $0.01$  &  \cite{pasachoff2015} \\
2015-03-20  &  $-0.79$  &  $1.61$  &  $0.12$  &  $0.23$  &  $0.01$  &  --  &  \\
2016-03-09  &  $-0.60$  &  $2.09$  &  $0.18$  &  $0.13$  &  $0.02$  &  --  &  \\
\hline
\end{tabular}

\end{table}

Compiled eclipse images that enclose the corona up to 2-3 solar radius (a bit inside the scope of SOHO/LASCO C2 field) enable the construction process of coronal flattening profile as a function of equatorial radius. Resulted profiles from 15 eclipse cases are presented in \autoref{fig3}, while shape parameters which consist of $\rmax$ and Ludendorff index are summarized in \autoref{tab2}. Compiled index from literatures are also presented as the main comparison together with solar activity phase as defined in \cite{ludendorff1928}, e.g. $\Phi=(T_{\text{ecl}}-T_{\text{min}})/|T_{\text{max}}-T_{\text{min}}|$, where $T_{\text{ecl}}$ is the time of eclipse, while $T_{\text{max}}$ and $T_{\text{min}}$ are the times of maximum and minimum of solar activity enclosing to $T_{\text{ecl}}$.

As shown in \autoref{fig3}, in most cases (80\% of the sample) flattening profile obviously indicate rise and fall that fit sufficiently good with the quadratic function. From this fitting, flattening indices reach maximum value at $\rmax$ that range between 1.2 to 2.1 {\rsun} with typical uncertainties below 0.2 \rsun. The largest value is $\rmax=2.09$ {\rsun} which was obtained from the last solar eclipse (9 March 2016). In this case, flattening profile shows a strong linear increase up to $\rmax$ and then sharply declines at larger radii. A rather flat brightness profiles at large radii produce larger uncertainty at this range. However, by employing statistical weighting, acceptable quadratic function of the flattening profile can be obtained.

There are two cases where the flattening profiles unfit significantly to quadratic functions and the implemented method failed to determine the reasonable value of $\rmax$. They are solar corona of 21 June 2001 and 13 November 2012 solar eclipses. Both eclipses occurred during the maximum phase of solar activity during which the Sun exhibited more helmet streamers. For the former case, streamers and coronal rays radiate from almost all over heliographic latitude such that solar corona appears to have a low flattening index. \cite{pishkalo2011} obtained Ludendorff index of $a+b=0.07$ from white-light portrait of 2001 solar corona while this study got a bit higher value, $a+b=0.11$. In general, there is declining trend of flattening index though fluctuations are observed and the values at the outer part have a larger uncertainties. If the examination is concentrated to the small radius, one may realize that the flattening profile reaches maximum approximately at $r=1.5$ \rsun, and than declines slightly.  In this case, statistical weighting is crucial and the obtained value of ${\rmax}\approx1.5$ {\rsun} and $a+b\approx0.11$ are sufficiently convincing.

Almost similar pattern is observed in the flattening profile of 2012 coronal image. Flattening index decreases monotonically to minimum allowed value such that $a+b=0.02$ was concluded. Polynomial fitting algorithm gave a parabolic curve directed upward since there are two data points with positive trend at large equatorial radius. This fitting is invalid. At small radius, flattening index can not be calculated because of saturated pixels. However, the produced flattening profile and determined Ludendorff index is in agreement with the result of \cite{pasachoff2015} who obtained $a+b=0.01$. Visual inspection may lead to approximated value of $\rmax=1.3$ \rsun since $\epsilon$ reach maximum at this radius.

\subsection{Shape Parameter Over Solar Cycles}

\begin{figure*}
\centering
\includegraphics[scale=0.8]{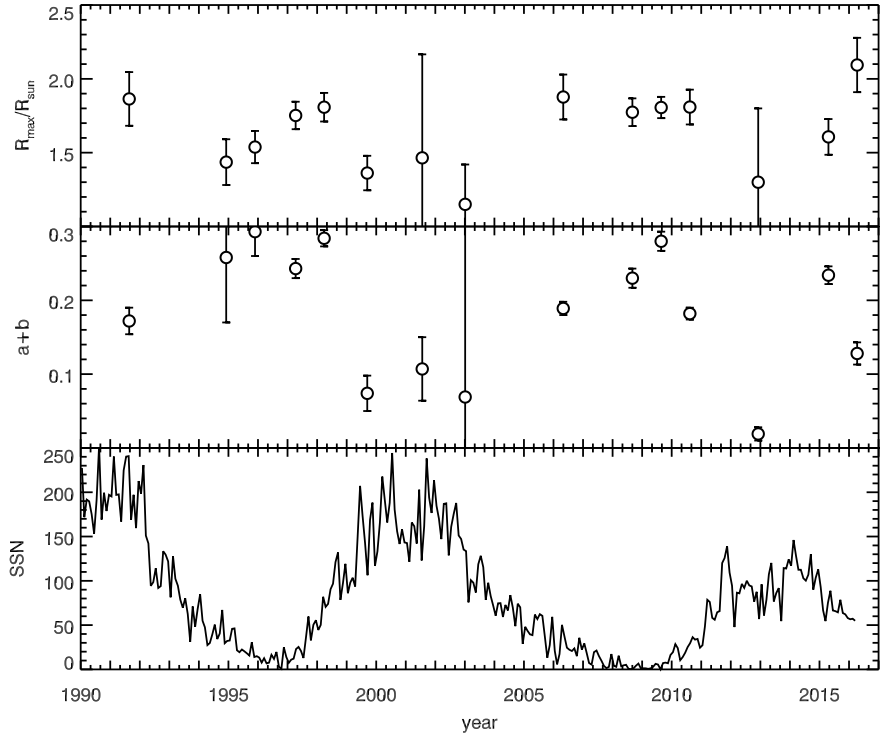}
\caption{Variation of shape parameters over solar activity cycle. Top panel displays $\rmax$ over time, while middle panel shows Ludedorff flattening index ($a+b$). Monthly average sunspot number ($SSN$) is plotted as comparison in bottom panel.}
\label{fig4}
\end{figure*}

\begin{figure*}
\centering
\includegraphics[scale=0.6]{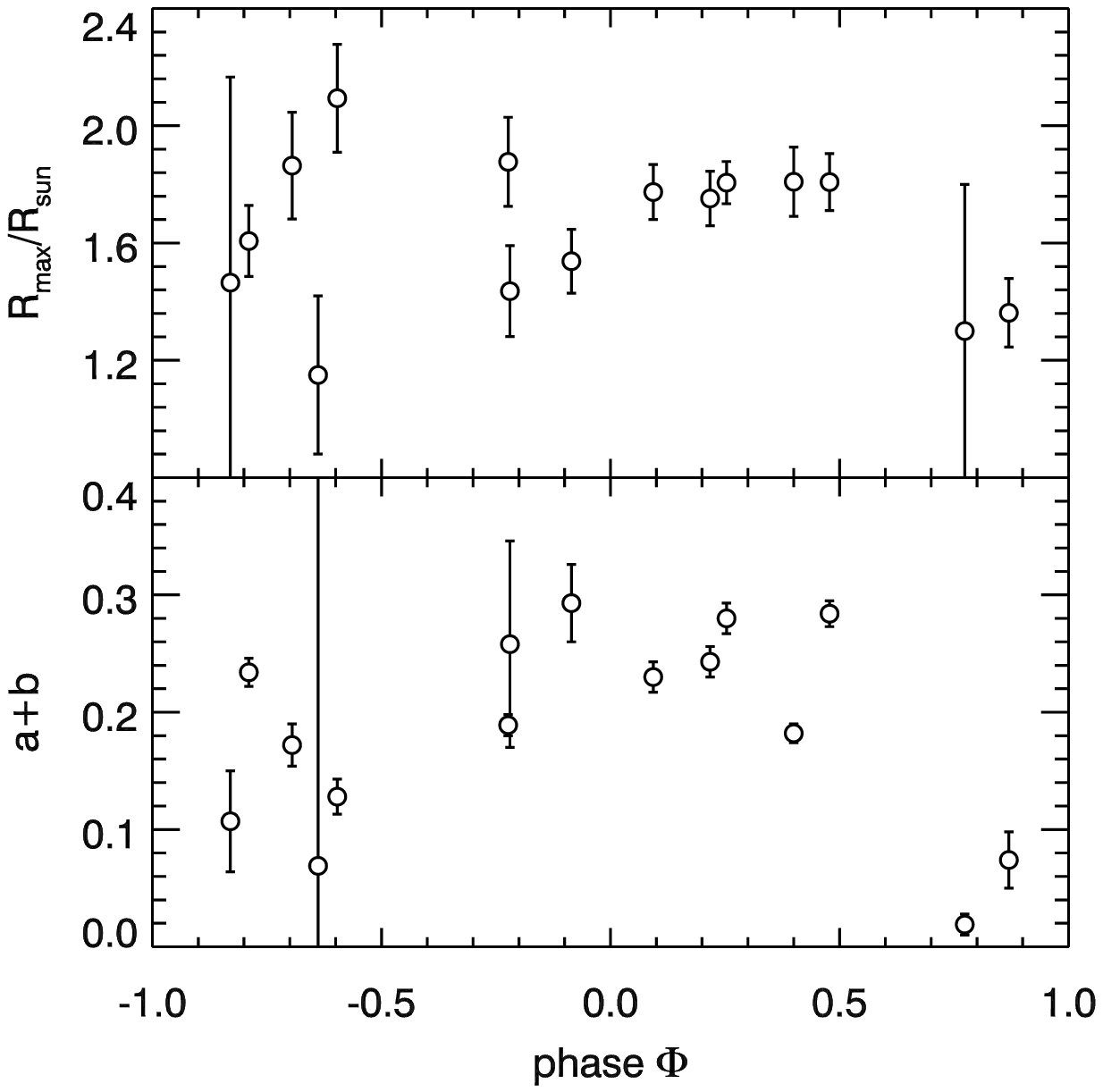}
\caption{Shape parameters as a function of phase $\Phi$ of solar activity.}
\label{fig5}
\end{figure*}

As summarized in \autoref{tab2}, Ludendorff flattening indices obtained from coronal images of 1991 to 2016 solar eclipses have values in the range of 0.02 to 0.29, while the radii of maximum flattening vary between 1.20 to 2.09 times solar radius. Both shape parameters change over time in the same phase of 11-years solar activity cycle (\autoref{fig4}). The shape parameters as a function of solar activity phase ($\Phi$) also presented in \autoref{fig5}. As reviewed by some authors \citep[e.g.][]{loucif1989,golub2009,pishkalo2011}, Ludendorff flattening index anti-correlates with sunspot number ($SSN$). During the minimum phase of solar activity ($\Phi\approx0$) observers tend to observe flattened corona with larger Ludendorff index and observe nearly circular corona during maximum ($|\Phi|\approx1$). Besides, the latitudinal extension of streamers-free polar regions becomes smaller during solar maximum \citep{loucif1989}. \autoref{fig4} also shows that $\rmax$ changes in similar way of flattening index.

Compared to the change of Ludendorff index, the cyclical variation of $\rmax$ is obscured by large dispersion at $\Phi\approx-0.7$. $\rmax$ that obtained from 1991, 2001, 2015, and 2016 can be regarded as deviating cases if 11-years cycle is expected. For 1991, the deviation is also occurred in Ludendorff index and might be related to the flattened corona that has been discussed by \cite{sykora1999}. They doubted the definition of Ludendorff flattening index with its regular change over solar cycle by arguing that the projected position of helmet structure on the celestial plane influences the appearance of the solar corona, which in turn, the observed shape and flattening. Due to the Carrington rotation for 2-3 days, the observed flattening index may change drastically. This can be considered as the source of intrinsic scatter of flattening index over solar activity cycle. For the case of 2001 solar eclipse, solar activity was at maximum level and helmet structures were distributed almost evenly in every direction (see \autoref{fig1}). The constructed flattening profile is somewhat flat with scatter at large radii. These conditions make the fitting a bit difficult. On the other side, the last two cases (2015 and 2016 eclipses) correspond to well-defined flattening profiles with relatively large $\rmax$ (see \autoref{fig3}).

The $\rmax$ value and its variation over solar cycle are rarely discussed in literatures. The following explanation is proposed to interpret this shape parameter and its change. It starts from the fact that the brightness of solar corona (K+F) in both equator and polar direction decline as the electron density drops exponentially \citep{newkirk1967,badalyan1996}. However, the declining rate in polar direction is a bit higher (steeper) compared to the one in equatorial direction \citep{lebecq1985,hanaoka2012}. While the coronal oblateness arise from the absolute difference of polar to equatorial brightness profile, the difference among declining rate causes variation of flattening index along solar distances. Flattening indices increase at $r<\rmax$ and decrease at $r>\rmax$. The value of $\rmax$ may change due to the change of brightness profile of the corona.

\cite{badalyan1996} examined white-light coronal images taken in 1952 to 1983 and found that density parameter of solar corona ($n_0$) in equator direction (assuming hydrostatic equilibrium) varies between $2\times10^8$ cm$^{-3}$ at minimum phase of solar cycle to $4\times10^8$ cm$^{-3}$ during maximum. Contrary, $n_0$ in polar direction fluctuates around $1\times10^8$ cm$^{-3}$ with insignificant amplitude over solar activity cycle. The discrepancy between equatorial change and polar change may explain the variation of $\rmax$ that depends on solar activity cycle.

\section{Conclusion}
\label{sec:conclusion}

In this study, 15 white-light solar coronal images taken during solar eclipses that occurred in 1991 to 2016 have been analyzed using semi-autonomous method such that shape parameter of the corona can be determined. Flattening profiles as a function of radius for very images have been produced. In most cases (80\% sample), the flattening profile can be modeled using 2nd order polynomial function such that the radius of maximum flattening ($\rmax$) can be determined. At small heliocentric distances ($r\leq\rmax$), flattening index increases almost linearly and the Ludendorff index (flattening at $r=2$ \rsun) can be extrapolated. In agreement with previous studies, Ludendorff index anti-correlates with monthly sunspot number. Additionally, this study shows that $\rmax$ change over solar cycle at the same phase of flattening index variation. The change of $\rmax$ can be interpreted as the observational consequences of the change in equatorial brightness profile that is different to the brightness profile in polar direction.

\normalem
\begin{acknowledgements}
The author thanks to the observers who share the coronal images that is analysed in this study. He also acknowledges Tiar Dani who introduced the study of solar corona structure, particularly for total solar eclipse of 9 March 2016.
\end{acknowledgements}

\bibliographystyle{raa}
\bibliography{ms0126}

\begin{thebibliography}{31}
\providecommand\natexlab[1]{#1}
\providecommand\JournalTitle[1]{#1}

\bibitem[Badalyan(1996)]{badalyan1996}
Badalyan, O. 1996, \aat, 9, 205

\bibitem[Badalyan \& S\'{y}kora(2008)]{badalyan2008}
Badalyan, O., \& S\'{y}kora, J. 2008, \coska, 38, 519

\bibitem[Bazin {et~al.}(2015)]{bazin2015}
Bazin, C., Vilinga, J., Wittich, R., {et~al.} 2015, in Proceedings of the
  Annual meeting of the French Society of Astronomy and Astrophysics, ed.
  M.~F., S.~Boissier, V.~Buat, L.~Cambr\'{e}sy, \& P.~Petit, 259

\bibitem[Dani {et~al.}(2016)]{dani2016}
Dani, T., Priyatikanto, R., \& Rachman, A. 2016, in International Symposium on
  Sun, Earth, and Life

\bibitem[Dorotovi\v{c} {et~al.}(1999)]{dorotovic1999}
Dorotovi\v{c}, I., Luka\v{c}, B., Minarovjech, M., {et~al.} 1999, \coska, 28,
  224

\bibitem[Espenak(1990)]{espenak1990}
Espenak, F. 1990, Fifty year canon of solar eclipses: 1986-2035, ed. {} (Sky
  Publishing Corporation)

\bibitem[Foukal(2004)]{foukal2004}
Foukal, P. 2004, Solar Astrophysics, Second Edition, ed. {} (Wiley-VCH)

\bibitem[Golub \& Pasachoff(2009)]{golub2009}
Golub, L., \& Pasachoff, J. 2009, The Solar Corona, 2nd edition, ed. {}
  (Cambridge University Press)

\bibitem[Gulyaev(1997)]{gulyaev1997}
Gulyaev, R. 1997, Astronomy \& Astrophysics Transactions, 13, 137

\bibitem[Hanaoka {et~al.}(2012)]{hanaoka2012}
Hanaoka, Y., Kikuta, Y., Nakazawa, J., Ohnishi, K., \& Shiota, K. 2012, \sophy,
  279, 75

\bibitem[Lebecq {et~al.}(1985)]{lebecq1985}
Lebecq, C., Koutchmy, S., \& Stellmacher, G. 1985, \aap, 152, 157

\bibitem[Loucif \& Koutchmy(1989)]{loucif1989}
Loucif, M., \& Koutchmy, S. 1989, \aaps, 77, 45

\bibitem[Ludendorff(1928)]{ludendorff1928}
Ludendorff, H. 1928, Sitzungsber. Preuss. Akad. Wiss. Phys.-Math. Kl., 16, 185

\bibitem[Newkirk(1967)]{newkirk1967}
Newkirk, G. 1967, \araa, 5, 213

\bibitem[Nikolsky(1956)]{nikolsky1956}
Nikolsky, G. 1956, Astron. Zh., 33, 84

\bibitem[Pasachoff {et~al.}(2007)]{pasachoff2007}
Pasachoff, J., Ru\v{s}in, V., Druckm\"{u}ller, M., \& Saniga, M. 2007, \apj,
  665, 824

\bibitem[Pasachoff {et~al.}(2009)]{pasachoff2009}
Pasachoff, J., Ru\v{s}in, V., Druckm\"{u}ller, M., Saniga, M., \& Minarovjech,
  M. 2009, \apj, 702, 1297

\bibitem[Pasachoff {et~al.}(2011{\natexlab{a}})]{pasachoff2011b}
Pasachoff, J., Ru\v{s}in, V., Druckm\"{u}llerova, H., {et~al.}
  2011{\natexlab{a}}, \apj, 734, 114

\bibitem[Pasachoff {et~al.}(2011{\natexlab{b}})]{pasachoff2011a}
Pasachoff, J., Ru\v{s}in, V., Saniga, M., Druckm\"{u}llerov\'{a}, H., \&
  Babcock, B. 2011{\natexlab{b}}, \apj, 742, 29

\bibitem[Pasachoff {et~al.}(2015)]{pasachoff2015}
Pasachoff, J., Ru\v{s}in, V., Saniga, M., {et~al.} 2015, \apj, 800, 90

\bibitem[Pint\'{e}r {et~al.}(1997)]{pinter1997}
Pint\'{e}r, T., Lorenc, M., Luka\v{c}, B., {et~al.} 1997, \coska, 27, 115

\bibitem[Pishkalo(2011)]{pishkalo2011}
Pishkalo, M. 2011, \sophy, 270, 347

\bibitem[Pishkalo \& Sadovenko(2008)]{pishkalo2008}
Pishkalo, M.~I., \& Sadovenko, E.~V. 2008, Kinematics and Physics of Celestial
  Bodies, 24, 44

\bibitem[Pishkalo \& Baransky(2009)]{pishkalo2009}
Pishkalo, N.~I., \& Baransky, A.~R. 2009, Kinematics and Physics of Celestial
  Bodies, 25, 315

\bibitem[Reginald {et~al.}(2003)]{reginald2003}
Reginald, N., {St. Cyr}, O., Davila, J., \& Brosius, J. 2003, \apj, 599, 596

\bibitem[Ru\v{s}in {et~al.}(1996)]{rusin1996}
Ru\v{s}in, V., Klocok, L., Minarovjech, M., \& Rybansk\'{y}, M. 1996, \coska,
  26, 37

\bibitem[Skomorovsky {et~al.}(2012)]{skomorovsky2012}
Skomorovsky, V., Trifonov, V., Mashnich, G., {et~al.} 2012, \sophy, 277, 267

\bibitem[Stoeva {et~al.}(2008)]{stoeva2008}
Stoeva, P., Stoev, A., Kuzin, S., {et~al.} 2008, \jastp, 70, 414

\bibitem[S\'{y}kora {et~al.}(2003)]{sykora2003}
S\'{y}kora, J., Badalyan, O., \& Obridko, V. 2003, \sophy, 212, 301

\bibitem[S\'{y}kora {et~al.}(1999)]{sykora1999}
S\'{y}kora, J., Badalyan, O., Obridko, V., \& Pint\'{e}r, T. 1999, \coska, 29,
  89

\bibitem[Tlatov(2010)]{tlatov2010}
Tlatov, A. 2010, \aap, 522, A27

\end{thebibliography}

\end{document}